\documentstyle[aps,12pt,psfig]{revtex}

\newcommand{ \be }{\begin{equation}}
\newcommand{ \ee }{\end{equation}}
\newcommand{ \bea }{\begin{eqnarray}}
\newcommand{ \eea }{\end{eqnarray}}
\newcommand{ \la }{\langle}
\newcommand{ \ra }{\rangle}

\newcommand{ \mpt }{\langle p_{t} \rangle}
\newcommand{ \phisubx }{\Phi_{x}}
\newcommand{ \sgmsqdyn }{\sigma^2_{\la x \ra,dynam}}
\newcommand{ \sgmsqincl }{\sigma^2_{x,incl}}
\newcommand{ \sgmincl }{\sigma_{x,incl}}

\newcommand{\f}{\frac}
\newcommand{\ovl}{\overline}

\begin{document}

\bigskip
\bigskip
\begin{flushright} {
LBNL-42992A \\
\it March 31, 1999
% \\ v. 11  === v2 at xxx
}

\end{flushright} 

\bigskip
\bigskip
\bigskip
\begin{center}
{\bf
Event-by-event fluctuations in collective quantities
} \\ 
\bigskip
\bigskip

S. A. Voloshin, V. Koch, H. G. Ritter
\\
\bigskip
{\em Nuclear Science Division,
Lawrence Berkeley National Laboratory,
\\   1 Cyclotron Rd., Berkeley, CA 94720}

\end{center}

\bigskip
\bigskip
\bigskip
{\footnotesize
\centerline{                ABSTRACT         }
\begin{quotation}

We discuss an event-by-event fluctuation analysis
of particle production in heavy ion collisions.
We compare different approaches to the evaluation of the event-by-event
dynamical fluctuations in quantities defined on groups of particles,
such quantities as mean transverse momentum, transverse momentum
spectra slope, strength of anisotropic flow, etc..
The direct computation of the dynamical fluctuations 
and the sub-event method are discussed in more detail. 
We also show how the fluctuation in different variables can be
related to each other.

\end{quotation}}
\bigskip
%
%
%%%%%%%%%%%%%%%%%%%%%%%%%%%%%%%%%%%%%%%%%%%%%%%%%%
%\newpage

\section{Introduction: Statistical, dynamical, and event-by-event fluctuations}

Any physical quantity measured in an experiment is subject to fluctuations.
In general, these fluctuations depend on the properties of the system
and may contain important information about that system.
In the context of heavy ion collisions, the system under consideration is a 
dense and hot fireball consisting of hadronic and/or possibly partonic matter.
The obvious challenge is to positively identify the
existence of a state of partonic matter early on in the life of the
fireball. 
The study of fluctuations may help in this task considerably.
First of all, fluctuations of a thermal system are directly related to
its  various susceptibilities~\cite{landau}, which in turn
are good indicators for possible phase changes.
For example, the extraction of the system heat capacity from
temperature  fluctuations has recently been proposed 
in~\cite{lstodol,lshur_ebe,lshur_qm}. 
Also, large event-by-event fluctuations
may indicate the existence of distinct event-classes, e.g. one with and one
without a Quark Gluon Plasma. 

Fluctuations have contributions of different nature.
First there are  `trivial' fluctuations due to a finite number of particles
used to define a particular observable in a given event. 
Examples of such observables are the  mean transverse momentum, $\mpt$,
where the average is taken over all particles in a given event, 
the strength of anisotropic flow, the ratios of multiplicities of different
particle species, etc.
Fluctuations due to finite multiplicity we shall call 
{\em statistical} fluctuations.
Statistical fluctuations can be evaluated by considering the production of all 
particles as totally independent.
All other fluctuations are of dynamical origin 
and shall be called {\em dynamical} fluctuations. 
Dynamical fluctuations can be subdivided into two classes: 
a) fluctuations
which do not change event-by-event (such as two-particle correlations
due to Bose-Einstein statistics or due to resonance decays), and 
b) fluctuations which occur on an event-by-event basis. 
The last ones we call event-by-event (EbE) fluctuations.
Examples of those are fluctuations in the ratio of charged to neutral
particle multiplicities due to creation of regions of DCC, 
or the fluctuations in anisotropic flow due
to creations of regions with ``unusually'' soft/hard equation of state. 
Also, the occurrence  of jets may give rise to event-by event 
fluctuations e.g. in the high $p_t$ tail of the transverse 
momentum distribution.   

The purpose of this paper is to present and discuss different approaches 
to the evaluation of dynamical fluctuations, in particular
 {\em EbE fluctuations}. We also address the 
limitations in extracting observables of physical relevance. 
Here we concentrate on fluctuations of the mean transverse momentum,
since experimental data for these fluctuations are already available 
\cite{lna49ebe}. Also,
fluctuations of the transverse momentum may be related to fluctuations of the 
temperature, which in turn may provide important information about 
the properties of the system under 
study~\cite{landau,lstodol,lshur_ebe,lshur_qm,lna49ebe}.  

In section II we present several  methods of
fluctuation analyses and illustrate them in terms of a simple
toy-model. We also discuss how these methods are related to each other
and to approaches already presented in the literature.
In the next sections we turn to the specific case of fluctuations of the
mean transverse momentum. We shall discuss the relation between fluctuations 
in $\mpt$ and the temperature. 
We finally address the question to what extent the heat
capacity of the system and the collision energy or centrality 
dependence thereof can be extracted from these fluctuations.

\section{Evaluation of fluctuations. ``Direct'' and ``sub-event''
methods}

\subsection{Definitions}

In this paper we consider fluctuations in collective quantities, 
the quantities defined on groups of particles. 
Such a group could be, for example, particles in some rapidity region.
It is useful to start with collective {\em average} (or intensive)
 quantities, which in rather general form can be defined as
\be
X \equiv \la x \ra =\f{\sum_i x_i }{M},
\label{ecv}
\ee
where $M$ is the particle multiplicity. 
The sum is taken over all particles in an event, 
and $x_i$ is a variable that is defined for each particle. 
For example, taking $x=p_t^2/(2m)$, where $p_t$ is the particle
transverse momentum and $m$ is the particle mass, 
would yield for $X$ an estimator for the (nonrelativistic)
temperature; taking $x=\cos(2(\phi-\Psi_{RP}))$, 
where $(\phi-\Psi_{RP})$ is the
particle azimuthal angle with respect to the reaction plane,
would give the strength of elliptic flow, $v_2$.
We use the same notation, $\la ... \ra$, for an average over 
all particles
in an event of a quantity defined on a particle, and also for an average
over all events of a quantity defined on an event. 
Then, $\la \mpt \ra$ would mean the average over all events of $\mpt$, 
the mean values of $p_t$ derived in each event. 
For an inclusive mean value (an average over the inclusive single particle
distribution) we reserve the notation $\ovl{x}$. 
For example, the inclusive mean transverse momentum 
we denote as $\ovl{p_t}$, which in
general does not necessarily coincide with $\la \mpt \ra $.

The fluctuations in quantity $X$ are defined by
\be
\sigma^2_X = \la X^2 \ra - \la X \ra ^2
= \la \la x \ra ^2\ra  - \la \la x \ra \ra ^2
\equiv\sigma^2_{X,stat}+\sigma^2_{X,dynam}.
\label{edf}
\ee
Note that $\sigma^2_{X,dynam}$ defined in this way 
can be negative provided that $\sigma^2_{X,stat}$ refers to
the statistical fluctuations in the totally ``uncorrelated'' particle 
production scenario, as defined above.
Dynamics (and/or kinematics) can suppress the fluctuations in
comparison to the case of the independent particle production.
Note, however, that the contribution to
$\sigma^2_{X,dynam}$ due to {\it event-by-event fluctuations} 
is always positive.

Within a given event sample
all three contributions to  $\sigma^2_{X}$, statistical fluctuations,
event-by-event fluctuations, and dynamical, not EbE, fluctuations, 
scale differently 
with the event multiplicity (see also~\cite{barsh,kadija}).
This property can be used as an additional criteria 
in the experimental separation of different
contributions~\cite{braune}.
Statistical fluctuations scale as  $\sigma^2_{X,stat} \propto 1/M$,
where $M$ is the event multiplicity. 
Event-by-event fluctuations, $\sigma^2_{X,EbE}$,  do not depend 
on multiplicity. 
The non-EbE dynamical fluctuations could have in general two terms, 
one which does not depend on multiplicity, and the second one, 
which similar to the statistical fluctuations scales 
as $ \propto 1/M$.
The part which does not depend on multiplicity is mainly due 
to Bose-Einstein correlations and two-particle final state interactions. 
We will argue below that the sub-event method
permits to eliminate this part from the total fluctuations.
The second part is due to resonance (jets) decays, momentum
conservation, etc..
Taking all facts together, it means that, provided the multiplicity 
independent part of dynamical non-EbE fluctuations is eliminated,
e.g. by the sub-event method, 
the multiplicity independent part of $\sigma^2_X$ is 
only of event-by-event origin.

\subsection{``Direct'' method}

Let us start with a simple example of a two dimensional 
nonrelativistic ideal gas in thermal equilibrium at temperature
$\tilde{T}$. 
In this case the particle  transverse momentum distribution is
\be
\f{dn}{dp_t^2} \propto \exp(-\f{p_t^2}{2m\tilde{T}}),
\ee
and the normalized probability density to find a particle with a given
$x\equiv p_t^2/(2m)$ is
\be
\f{dw}{dx} = \f{1}{\tilde{T}} \exp (\f{-x}{\tilde{T}}).
\label{etd}
\ee
Each event in this example is a random selection of $M$
particles from a thermal bath.
For simplicity we assume that  $M$ is constant\footnote{
It is sufficient here to assume that the distribution in $x$
does not depend on the event multiplicity. 
If this is not the case,
it would mean a known source of event-by-event fluctuations
(fluctuations in multiplicity), which in principle should 
be analyzed separately. See also~\cite{bk99,fr99} for more detail
treatment of the multiplicity fluctuations.}.

In order to get an estimate for the temperature of the system
one needs to fit the slope of the $p_t$ distribution.
Application of a maximum likelihood method yields
the best estimator for $T$
\be
T=\frac{\sum_i x_i}{M} 
\equiv \la x \ra 
=\la \f{p_t^2}{2m} \ra,
\ee
which is just the result of the equipartition theorem in two dimensions
($\la E \ra   /M = 2 \cdot (T/2)$).  

The statistical fluctuations in the quantity $T$ (the
fluctuations due to finite multiplicity $M$, under assumption of 
the independent particle production) can be directly calculated:
\be
\sigma_{T,stat}^2=\la T^2 \ra -\la T \ra^2 
=\la (\frac{\sum_i x_i}{M})^2 \ra - \la \frac{\sum_i x_i}{M} \ra^2
=\f{\sigma_{x,incl}^2}{M},
\ee
where $\sigma_{x,incl}^2\equiv \overline{(x-\overline{x})^2}$ 
is the variance of the inclusive distribution in $x$. 
For a thermal distribution~(\ref{etd}) one has $\overline{x}=\tilde{T}$, 
$\overline{x^2} =2\tilde{T}^2$, and $\sigma^2_{T,stat}=\tilde{T}^2$.
It follows that 
\be
\f{(\Delta T)^2_{stat}}{T^2} 
\equiv \f{\sigma_{T,stat}^2}{\tilde{T}^2} 
= \f{\sigma_{x,incl}^2}{M\tilde{T}^2}=\f{1}{M}.
\label{edeltaTstat}
\ee

For practical applications it is very important to know the accuracy
in the calculation of $\sigma_T$. 
The direct calculation of the variance of $\sigma_T$,
$\sigma^2_{\sigma^2_T}$ is straightforward, but rather lengthy.
For a large number of events ($N_{ev} \gg 1$) the answer is 
simple~\cite{pdb}:
\be
\sigma^2_{\sigma_T^2} \approx \f{2\sigma^4_T}{N_{ev}} 
\Longrightarrow
\sigma^2_{\sigma_T} \approx \f{\sigma^2_T}{2N_{ev}} 
\Longrightarrow
\sigma_{\sigma_T} \approx \f{\sigma_T}{\sqrt{2N_{ev}}}.
\ee
In this paper we consider mostly the case when statistical
fluctuations are much larger than the dynamical ones.
In this case $\sigma_T \approx \sigma_{T,stat}$ and
\be
\sigma_{\sigma_T} 
\approx \f{\sigma_{T,stat}}{\sqrt{2N_{ev}}}
= \f{1}{\sqrt{2N_{ev} M}}.
\ee

\subsection{``Sub-event method''}

It is simpler to use the sub-event method for the
calculation of the EbE fluctuations. 
Just recollect that we are interested in the fluctuations which affect the
entire event. 
If one subdivides such an event into two subsystems, which we
call below sub-events ``a'' and ``b'', the quantities defined on these two 
sub-events should be correlated.
\bea
\la \, (T_a - \la T_a \ra) \, (T_b  - \la T_b \ra) \, \ra &=&
\la \, ((\Delta T_a)_{stat}+(\Delta T)_{dynam}) 
((\Delta T_b)_{stat}+(\Delta T)_{dynam}) \, \ra 
\nonumber
\\
&=& \sigma^2_{T,dynam}.
\label{ecorr}
\eea
Note that in the case of an ideal gas, where the fluctuations are entirely
of  statistical nature the above correlator would yield zero.

The sub-event method permits one to avoid some problems
of the ``direct'' computation of EbE fluctuations.
In particular the problems
related to the separation of the EbE fluctuations from other dynamical
effects, such as Bose-Einstein correlations (the HBT effect). 
It is not possible to avoid the HBT correlations in the direct
approach and one can only perform a rather complicated estimate 
of its contribution (see, for example,~\cite{lna49ebe}). 
In the sub-event method one can define the sub-events on different 
regions, so that particles from two regions are not correlated (in the
HBT sense), and the problem simply disappears.
For example, one can define sub-events on rapidity regions separated 
by  0.1 unit of rapidity.
The same trick can be used to get rid of the ``two track resolution''
problem which is quite serious in many experiments.
In addition, using the sub-event method it is also possible to study how 
the ``proton temperature'' is correlated
with the ``pion temperature'' and many other interesting questions.
Unfortunately, we do not know any simple way of avoiding
the correlations due to energy-momentum conservation (see also the
discussion of this question in~\cite{lposk,odyniec}).

Another way to look at the correlations using the sub-event method is to
compare widths of the distributions in $(T_a-T_b)$ and in $(T_a+T_b)$.
While the first distribution is governed mostly by statistical
fluctuations, the second one contains dynamical fluctuations as well. 
The difference in the width of the distributions would yield 
the dynamical fluctuations (see the calculations within the toy model below 
in this section).

\subsection{Relations to other methods}

The function $\phisubx$ is frequently used in the
literature~\cite{lgazd,lbleicher,lmrov_compr,lltgs,lna49ebe}
for the event-by-event fluctuations study. 
It is defined as
\be
\Phi_{x}=\sqrt{\la Z^2 \ra /\la M \ra} - \sqrt{\overline{z^2}},
\label{ephisubx}
\ee 
where 
\be
Z=\sum_i z_i, \;\;\; z_i=x-\overline{x} ,
\ee
and $x$ is the quantity under study, for example, the transverse momentum.
In order to compare  $\phisubx$ and $\sgmsqdyn$ results, 
we first derive a useful formula. 
We start with the definition of $\phisubx$, given by Eq.~(\ref{ephisubx}).
Multiplying both sides of the equation by 
$(\sqrt{\la Z^2 \ra /\la M \ra} + \sqrt{\ovl{z^2}})$,
and taking into account that
$\phisubx$ is the difference  between two almost equal quantities,  
($\sqrt{\la Z^2 \ra /\la M \ra} \approx \sqrt{\overline{z^2}}
\equiv \sigma_{x,incl}$)
one gets
\bea
2 \phisubx \sgmincl &\approx& 
\Phi_{x}(\sqrt{\la Z^2 \ra /\la M \ra} +\sgmincl)
\nonumber
\\
&=&
(\sqrt{\la Z^2 \ra /\la M \ra} -\sgmincl) 
(\sqrt{\la Z^2 \ra /\la M \ra} +\sgmincl)
\nonumber
\\
&=&
\la Z^2 \ra /\la M \ra -\sgmsqincl.
\eea
To proceed further we need the assumption that multiplicity is not
correlated with the distributions in $x$.
Under this assumption
\bea
\f{\la Z^2 \ra }{\la M \ra} -\sgmsqincl
&=&
\f{\la M \ra \ovl{(x-\overline{x})^2} +
\la M(M-1) \ra \la (x_{i} -\overline{x})(x_{j} -\overline{x}) \ra }
{\la M \ra}
-\sgmsqincl
\nonumber \\
&=&
\f{\la M (M-1)\ra}{\la M \ra}  \la (x_{i} -\overline{x})
(x_{j} -\overline{x}) \ra
\nonumber \\  
&\approx& 
\la M \ra \sgmsqdyn
\eea
We end up with the formula (see also~\cite{trainor}):
\be
\phisubx \approx \f{\sgmsqdyn \la M \ra}{2\sgmincl}.
\ee
From this formula one can see both, strong and weak points of the two 
analyses using $\phisubx$ and $\sgmsqdyn$.
The quantity which is directly related to the underlying physics
is $\sgmsqdyn$. In this sense it is preferable.  
On the other hand, if one want to compare different systems in order
to see if the underlying physics is the same, 
and events (systems) differ only by the total multiplicity, 
then one has to take into account that the correlations scale
inversely proportional to the event  multiplicity. 
In this sense the multiplication of  $\sgmsqdyn$ by (in this case, observed) 
multiplicity allows one to check if the physics is changing.
This is the advantage of the  $\phisubx$ approach (as well as any
other approach dealing with the quantity proportional to
$\sgmsqdyn \la M \ra$).
But one should be careful when comparing $\phisubx$ measured by
different experiments, and even by the same experiment but under
different conditions and/or analysis cuts. 
$\phisubx$ is scaled by the {\em observed} multiplicity. 
It means that even for the same event sample
it would depend, for example, on the track selection cuts.

It is clear from the definition~(\ref{ecv}) that correlations between 
the average collective quantities ($\la X_a X_b \ra$) and the corresponding
fluctuations (in other words, autocorrelations, $\la X_a X_a \ra$) 
can be rewritten using the usual two-particle correlations (the same, as, 
for example, the second factorial moment used in the study of 
intermittency, or the discussed above quantity $\Phi_x$). 
In this sense the correlations in collective variables provide no
additional information compared to the two-particle correlations.
Details and subtleties of the relation between
the two particle correlations and the even-by-event observables 
have been discussed recently in \cite{bk99,fr99}. 

It should be noted, on the other hand, that it can be much more
convenient to work with collective variables.
The ``signal-to-background'' ratio,
i.e. $\sigma_{dynam}/\sigma_{stat}$, 
in these variables generally grows as
$\sqrt{M}$, where $M$ is the multiplicity.
The reason for this is that fluctuations in ``background''
distribution scale as $\sqrt{M}$ while the ``signal'' (strength of
flow, change in $p_t$, ...) would depend linearly on multiplicity.
A good ``signal-to-background'' ratio can be very
important in order to select ``unusual'' events, i.e. the events with
particular strong/weak signal (temperature, strength of flow, etc.).
Another advantage of using the quantities defined on a group of
particles is a practical one related to computing time. 
The computation of the two particle correlation function 
in the traditional way using events with multiplicity of a few hundred 
or even a few thousand particles does require a lot of computing time.

\subsection{A toy model}

Let us conclude this section by employing a 
toy Monte-Carlo event generator  in order to illustrate
how the above discussed formulae work. 
In this toy model we generate a few sets of 4000 events each; 
all events are of the same multiplicity $M=1000$. 
The different sets are generated for different event-by-event
fluctuations in temperature, which is distributed in accordance with
\be
T=\tilde{T} (1+\delta (r-0.5)),
\ee
where $r$ is a random number in [0,1], and
$\delta$ is a parameter responsible for the strength of the fluctuations.
The transverse momentum of each particle is generated in accordance
to the distribution  
\be
\f{dn}{dp_x dp_y} \propto \exp(-\f{p_x^2+p_y^2}{2mT}).
\ee

Using the data generated for $\tilde{T}=0.05$~GeV and $\delta=0.03$ 
and $\delta=0.1$ (the first one is close to the limit of our 
sensitivity to the dynamical fluctuations
for such a data set) we calculate the dynamical fluctuations (in
accordance to Eq.~(\ref{edf})) for each group of 500 events.
The results are presented in Fig.~1 together with a fit to a constant.
The fit values should be compared to the input values of
$\sigma^2_{dynam}/\tilde{T}^2=\delta^2/12=(0.03)^2/12=0.75\cdot 10^{-4}$,
and  $(0.1)^2/12=0.833\cdot 10^{-3}$ respectively.  
A good agreement between the input and the reconstructed values is
observed.
It is remarkable that the method is sensitive to fluctuations
which one would not expect judging only from the single particle spectrum.
The distribution in $p_t^2$ for the case of $\delta=0.1$ is presented
in Fig.~2 together with an exponential fit. 
Not only is no deviation from an exponential distribution visible, 
but the fit quality is very
good, $\chi^2/n.d.f. =57/72$.

The next figures, Figures~3--5, are for illustration of the sub-event method.
Fig.~3 shows the correlation between temperatures measured in two
sub-events.
Already from the scatter plot one can see that the two quantities are
correlated, which is the consequence of the introduced event-by-event
fluctuations. 
The profile plot, which shows the average temperature of
the subevent ``b'' as a function of the temperature observed in the
sub-event ``a'', looks even more convincing.
One can see that the temperature values
reconstructed on two different subevents are closely correlated. 
Such an observation unambiguously indicates a presence of dynamical
correlations in the data. 

Another way to study if the temperature values are correlated is to
look at the distributions in  $(T_a - T_b)$ and $(T_a + T_b)$, as
discussed above. These distributions are presented in Fig.~4.
One can see that the distribution in  $(T_a - T_b)$, 
containing only statistical fluctuations, is significantly narrower
than the distribution in  $(T_a + T_b)$, which has both statistical and
dynamical fluctuations. 
Using just the RMS values from the plots, one can estimate the
dynamical fluctuations as $\sigma_{T,dynam}^2=((4.33)^2-(3.21)^2)\cdot
10^{-6}/4=0.211\cdot 10^{-5}$~GeV$^2$.

Quantitative analysis  of the dynamical fluctuations 
using Eq.~(\ref{ecorr}) is presented in Fig.~5.
The observed strength of the correlation 
$\sigma_{T,dynam}^2=(0.205\pm0.011) \cdot 10^{-5}$~GeV$^2$
 should be compared with the input value of 
$\sigma_{T,dynam}^2 = (\tilde{T} \delta)^2/12=
(0.05 \cdot 0.1)^2/12=0.208\cdot 10^{-5}$~GeV$^2$.

\section{Correlations between different collective variables. Relations
between fluctuations}

Often the fluctuations in different variables are tightly
connected with each other.
For example, let us consider fluctuations in the mean transverse
momentum $\mpt$ and fluctuations in the effective temperature (more 
precisely, in the slope parameter of the transverse momentum 
distribution) $T$.
We assume that  $\mpt$ is uniquely defined by this parameter.
Then one can write $\la \mpt \ra = F(\la T \ra)$.
Assuming that the fluctuations are of Gaussian nature, 
arguments from the theory of error propagation  give: 
\be
\sigma_{\mpt,dynam} = |F'(\la T \ra)| \; \sigma_{T, dynam} 
\Longrightarrow
\f{\sigma_{\mpt,dynam}}{ \la \mpt \ra } 
= |\f{F'(\la T \ra )}{F(\la T \ra)}| \; \sigma_{T,dynam}. 
\ee

In reality, the $p_t$  spectra of most particles lie in between two
limiting cases $\la \mpt \ra \propto \sqrt{\la T \ra} $
(nonrelativistic ideal gas) and
 $\la \mpt \ra \propto \la T \ra $ (ultrarelativistic ideal gas).
It follows then that 
\be
\f{\sigma_{\mpt, dynam}}{\la \mpt \ra} 
= (0.5 \div 1)  \f{\sigma_T}{\la T \ra}. 
\ee
One can apply this relation to recent measurements~\cite{lna49ebe}.
In this paper the limits on EbE fluctuation of $\mpt$ 
was established as $\sigma_{\mpt}/\la \mpt \ra <0.01$.
According to our conclusion it means that $\sigma_T/\la T \ra < 0.02$
(conservative estimate). 
Note that the mean multiplicity used in this
experiment is of the order of $\la M \ra \approx 250$ and the 
statistical fluctuations in the temperature are of the order of 
 $\sigma_{T,stat}/\la T \ra  \approx 1/\sqrt{\la M \ra} \approx 0.07$.

The relation between effective temperature and mean transverse
momentum becomes less transparent if at the time of thermal freeze-out
sizeable energy/momentum dependent mean field potentials are present. 
This could be due to mass changes as proposed in the context of chiral
symmetry restoration or simply due to long range interactions among the
particles. In this case 
the relation between transverse momentum and temperature, $F(T)$, 
depends on the detailed structure of the mean field forces at play.

\section{Can we really measure $C_V$ using $p_t$ spectra?}

It has been proposed in~\cite{lstodol,lshur_ebe,lshur_qm} to measure
temperature fluctuations in order to access the heat capacity of the
system
\be
(\f{\Delta T}{T})^2 \equiv \f{\sigma_T}{\la T \ra} = \f{1}{C_V}.
\label{ehc}
\ee
Such measurements, if possible, can provide very important
information about the equation of state, and can be used to detect 
the phase transitions where the heat capacity could undergo very rapid change.
The possibility to get such information becomes
one of the major attractions of event-by-event physics.
It was assumed in~\cite{lstodol,lshur_ebe} that the temperature 
fluctuations can be evaluated 
using an event-by-event analysis of the transverse momentum spectra.
In this section we question this particular possibility.
% mainly for practical reasons.  
Our conclusion is that 
1) the required temperature fluctuations {\em cannot}
be measured using the information on only particle transverse
momentum, and 
2) even if the transverse spectra slope fluctuations are
sensitive to phase transition, such a relation is more
complicated then suggested by Eq.~(\ref{ehc}).

Our arguments are based on the following observations. 
Let us consider a two dimensional ideal gas at temperature $\tilde{T}$.
We would like to use $M$ particles to define the temperature 
by measuring $p_t$ spectra. 
For simplicity, $M$ is fixed.
An estimate of the temperature would be 
\be
T=\f{\sum (p_x^2+p_y^2)/(2m)}{M}, \;\;\; \la T \ra = \tilde{T}.
\ee
The event-by-event fluctuations in $T$ can be easily estimated. They are
\be
(\f{\sigma_T}{\la T \ra})^2=(\f{\sigma_{T,stat}}{\la T \ra})^2
=\f{1}{M} =\f{1}{C_V},
\ee
taking into account that the heat capacity of a system of $M$ particles of
a two dimensional ideal gas is $C_V=2 \cdot (M/2)=M$.
This formula coincides with Eq.~(\ref{ehc}).
Now let us take a three dimensional ideal gas, but use only
two components of the particle momentum ($p_x$ and $p_y$) for an estimate 
of the temperature.
It is obvious that the fluctuations in $T$ quantitatively do not 
change compared to the two dimensional case, but 
now they clearly do not provide us with the knowledge of the heat capacity.
The heat capacity has changed to $C_V=3/2 \, M$.
One can continue with such arguments adding to the consideration
internal degrees of freedom: the observed fluctuations remain the same
while the heat capacity continues to change.
Thus,  our conclusion on the possibility to access the system heat
capacity by measuring the fluctuations in transverse momentum slopes
are rather pessimistic.
However, if the fraction of the heat capacity that actually is being
measured remains constant, one could still hope to see rapid
changes in that quantity as the system goes through a phase
transition. So, it is definitely interesting to measure
an excitation function of the mean transverse momentum fluctuations.

\section{Summary}

We have studied event-by-event fluctuations with 
the direct method and have introduced a new way to determine
fluctuations with the sub-event method. A suitable choice of
sub-events and the possible combination of particles within a
sub-event or between sub-events trivially allows to exclude some 
dynamical correlations like the HBT correlations or experimental 
effects like two particle resolution effects.

The relationship to the $\Phi_{x}$ variable has been discussed. 
The fact that correlations between different collective quantities 
and their fluctuations can be formulated in terms 
of two-particle correlations has also been discussed in other 
papers~\cite{bk99}. 
The importance of the signal-to-background ratio has been pointed 
out and the fact that large multiplicity detectors help 
to increase this ratio.

We have applied the methods developed to a toy model and find that 
fluctuations can be determined with very high sensitivity.

It has been proposed to measure the heat capacity of a system by
studying the dynamical temperature fluctuations. We have shown, that
the heat capacity cannot be measured from the temperature
fluctuations. However, it cannot be excluded that by carefully
measuring an excitation function and the related fluctuations a
possible phase transition would manifest itself in increased
fluctuations in a (narrow) energy region.

%\newpage
\section*{Acknowledgments}

We are grateful to A. Poskanzer, G. Rai, and other members of the RNC
group for many useful discussions.

This work was supported by the Director, Office of Energy Research,
Office of High Energy and Nuclear Physics, Division of Nuclear Physics
of the U.S. Department of Energy under Contract DE-AC03-76SF00098.

\clearpage
\newpage
\section*{Figure captions}

\begin{enumerate}
\item
        Reconstructed dynamical fluctuations for $\delta=0.03$ (left
panel ) and $\delta=0.1$ (right panel).

\item
        Particle distribution in $p_t^2$ together with an exponential fit.

\item
        Sub-event method. Correlations between $T_a$ and $T_b$. Scatter (left
panel) and profile (right panel) plots.

\item
        Sub-event method. Distribution in  $T_a - T_b$ (left
panel) and $T_a + T_b$ (right panel).
 
\item
        Sub-event method. 
$\la (T_a -\la T_a \ra)(T_b -\la T_b \ra)\ra \equiv
\sigma_{T,dynam}^2$ calculated on the 500 event subsamples
for $\delta=0.1$.

\end{enumerate}

\clearpage
\newpage
\begin{figure}
\centerline{\psfig{figure=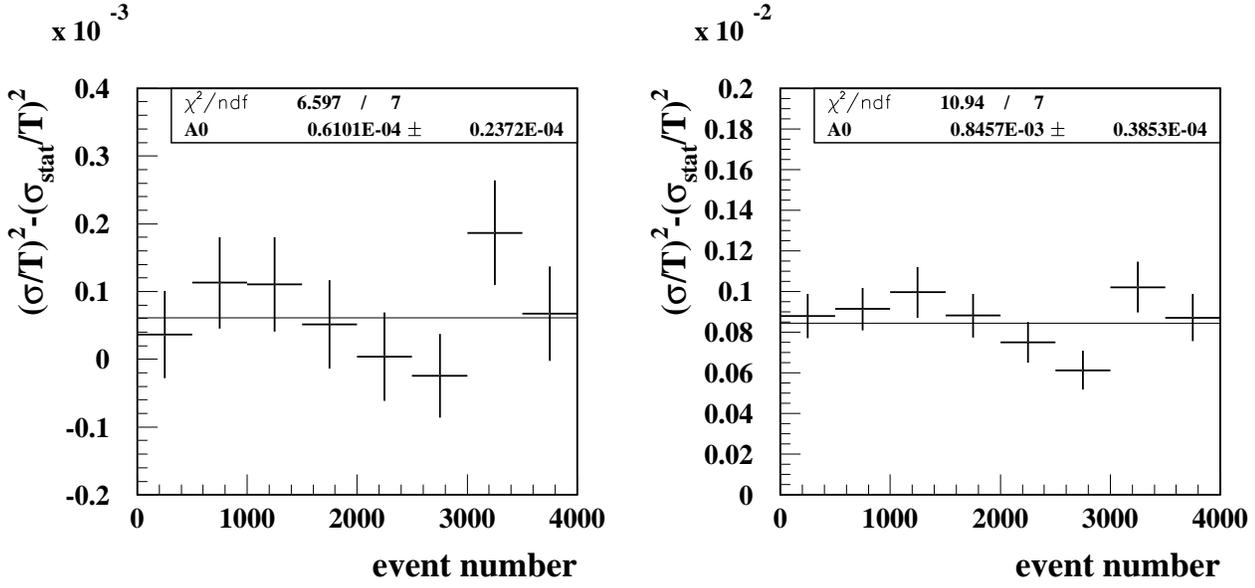,height=9.0cm}}
  \caption[]{
        Reconstructed dynamical fluctuations for $\delta=0.03$ (left
panel ) and $\delta=0.1$ (right panel).
    }
\label{fig1}
\end{figure}

\begin{figure}
\centerline{\psfig{figure=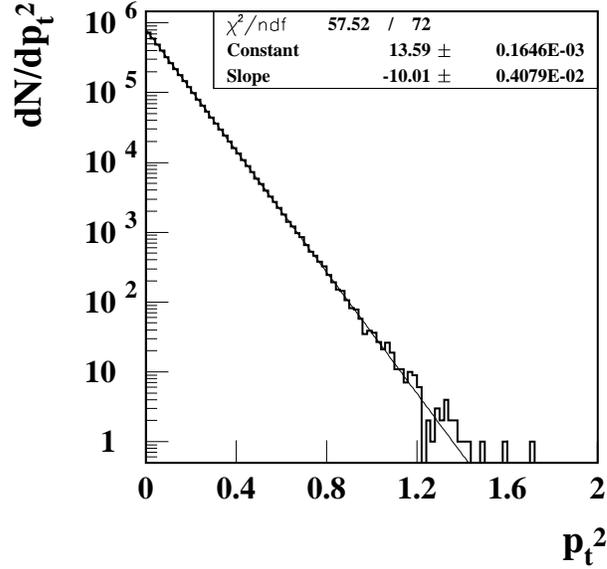,height=10.0cm}}
  \caption[]{
        Particle distribution in $p_t^2$ together with an exponential fit.
    }
\label{fig2}
\end{figure}

\begin{figure}
\centerline{\psfig{figure=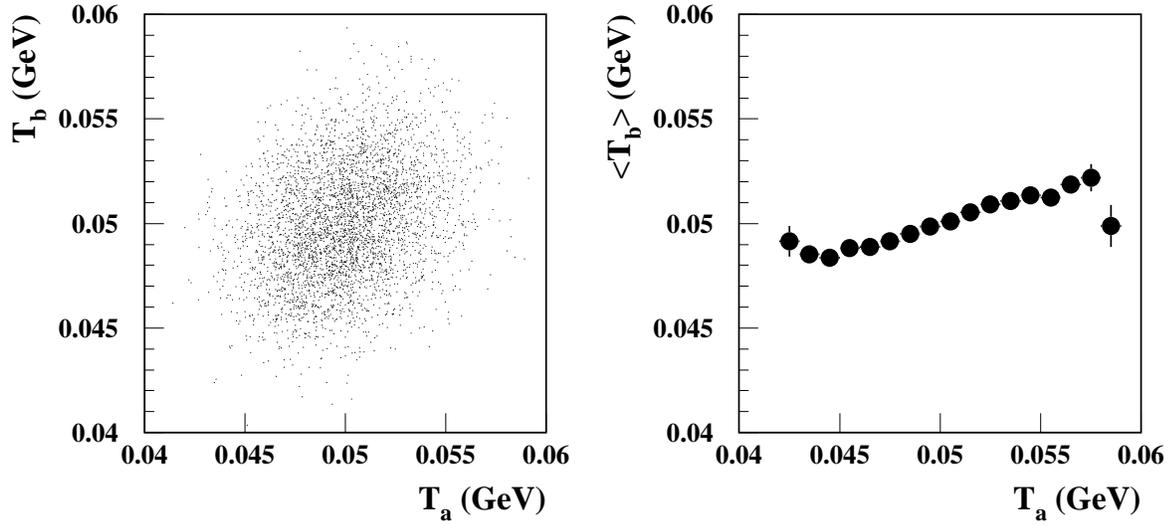,height=9.0cm}}
  \caption[]{
        Sub-event method. Correlations between $T_a$ and $T_b$. Scatter (left
panel) and profile (right panel) plots.
    }
\label{fig3}
\end{figure}

\begin{figure}
\centerline{\psfig{figure=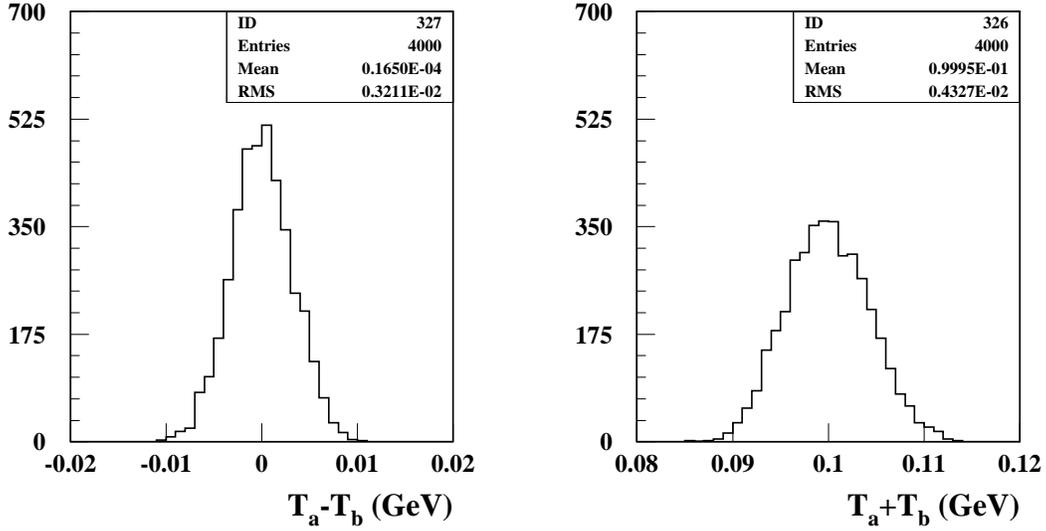,height=9.0cm}}
  \caption[]{
        Sub-event method. Distribution in  $T_a - T_b$ (left
panel) and $T_a + T_b$ (right panel).

    }
\label{fig4}
\end{figure}

\begin{figure}
\centerline{\psfig{figure=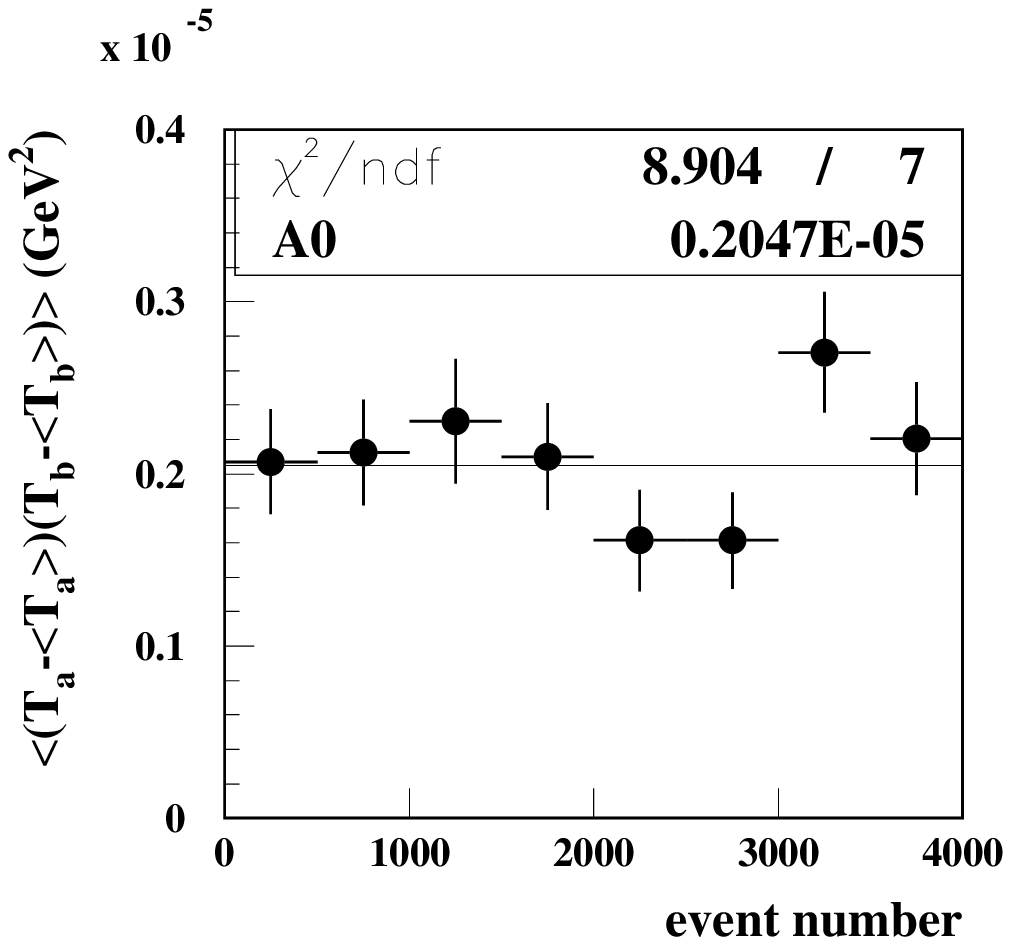,height=9.0cm}}
  \caption[]{
        Sub-event method. 
$\la (T_a -\la T_a \ra)(T_b -\la T_b \ra)\ra \equiv
\sigma_{T,dynam}^2$ calculated on the 500 event subsamples
for $\delta=0.1$.
    }
\label{fig5}
\end{figure}


\begin{references}

\bibitem{landau} L.L. Landau and E.M.. Lifshitz, {\em Statistical
Physics}, Pergamon Press, 1958.

\bibitem{lstodol} L. Stodolsky, Phys. Rev. Lett., 75 (1995) 1044.

\bibitem{lshur_ebe} E.V. Shuryak, Phys. Lett., B423 (1998) 9.

\bibitem{lshur_qm} E.V. Shuryak, Nucl. Phys., A638 (1998) 207c.

\bibitem{lna49ebe} G. Roland for the NA49 Collaboration,
Nucl. Phys. A638 (1998) 91c;
H. Appelshauser et al., NA49 Collaboration, to be published.

\bibitem{barsh} S. Barshay, H. Braun, J.P. Gerber,
and G. Maurer, Phys. Rev., D21 (1980) 1849; 
S. Barshay, Phys. Rev. D29 (1984) 1010.

\bibitem{kadija} K. Kadija and M. Martinis, Z. Phys., C56 (1992) 437.

\bibitem{braune} S. Braune et al., Phys. Lett., 123B (1983) 467.

\bibitem{bk99} A. Bialas and V. Koch, preprint nucl-th/9902063, 1999.

\bibitem{fr99} M. Belkacem et al., preprint nucl-th/9903017, 1999.
 
\bibitem{pdb} Review of Particle Physics, R. M. Barnett et al., 
Phys.Rev., D54 (1996). 

\bibitem{lposk} A.M. Poskanzer and S.A. Voloshin, Phys. Rev., C53
(1998) 896. 

\bibitem{odyniec} G. Odyniec, preprint nucl-ex/9901001, 1999.

\bibitem{lgazd} M. Gazdzicki and S. Mrowczynski, Z.Phys., C54 (1992) 127.  

\bibitem{lbleicher} M. Bleicher et al., Phys. Lett., B435 (1998) 9.

\bibitem{lmrov_compr} S. Mrowczynski, Phys. Lett., B430 (1998) 9.

\bibitem{lltgs} F. Liu, A. Tai, M. Gazdzicki, and R. Stock,
preprint hep-ph/9809320, 1998.

\bibitem{trainor} T.A. Trainor, to be published in Proceedings of the
15th Winter Workshop on Nuclear Dynamics, Park City, UT, January
9--16, 1999, Kluwer Academic Press.


\end{references}
\end{document}